\documentclass{llncs}
\usepackage{amsmath}
\usepackage{amsfonts}
\usepackage{amssymb}
\usepackage{stmaryrd}

\newcommand{\SROIQ}{$\mathcal{SROIQ}$}

\newcommand{\PSpace}{{\sc{PSpace}}}
\newcommand{\NPSpace}{{\sc{NPSpace}}}
\newcommand{\PTime}{{\sc{P}}}
\newcommand{\NExpTime}{{\sc{NExpTime}}}

\newcommand{\Inter}{\mathcal{I}} 
\newcommand{\true}{\mathit{true}}
\newcommand{\false}{\mathit{false}}

\RequirePackage{graphicx}

\newcommand{\atleast}[1]{\mathord{\geqslant}#1\,}
\newcommand{\atmost}[1]{\mathord{\leqslant}#1\,}

\newcommand{\con}[1]{#1}
\newcommand{\rol}[1]{#1}

\newcommand{\rolR}{\rol{r}}
\newcommand{\rolS}{\rol{s}}

\newcommand{\rolU}{\rol{u}}

\newcommand{\conC}{\con{C}}
\newcommand{\conD}{\con{D}}

\usepackage{xspace}
\usepackage{pgf,tikz}
\usepackage{pgfplots}
\usepackage{url}

\newcommand{\wrt}{w.r.t.\ }
\newcommand{\ifft}{iff\xspace}
\newcommand{\ie}{i.e.\ }
\newcommand{\eg}{e.g.\ }

\newcommand{\cf}{cf.\ }

\newcommand{\NP}{\mbox{\sc{NP}}\xspace}

\newcommand{\ExpTime}{\mbox{\sc{ExpTime}}\xspace}
\newcommand{\TwoExpTime}{\mbox{\sc{2\ExpTime}}\xspace}
\newcommand{\NTwoExpTime}{\mbox{\sc{N\TwoExpTime}}\xspace}

\newcommand{\sroiq}{\text{$\mathcal{SROIQ}$}\xspace}

\newcommand{\KB}{\text{$\mathcal{K}$}\xspace}

\newcommand{\Dis}{\text{$\mathsf{Dis}$}\xspace}
\newcommand{\Self}{\ensuremath{\mathit{Self}}}
\newcommand{\concept}[1]{\ensuremath{\mathsf{#1}}\xspace}
\newcommand{\role}[1]{\ensuremath{\mathsf{#1}}\xspace}

\newcommand{\NcDL}{\text{$N_{C}$}}
\newcommand{\NrDL}{\text{$N_{R}$}}
\newcommand{\NiDL}{\text{$N_{I}$}}

\newcommand{\calR}{\text{$\mathcal{R}$}\xspace}
\newcommand{\Rh}{\text{$\mathcal{R}_h$}\xspace}
\newcommand{\Ra}{\text{$\mathcal{R}_a$}\xspace}

\newcommand{\bfC}{\text{$\mathbf{C}$}\xspace}
\newcommand{\calT}{\text{$\mathcal{T}$}\xspace}
\newcommand{\calA}{\text{$\mathcal{A}$}\xspace}
\newcommand{\calC}{\ensuremath{\mathcal{C}}\xspace}
\newcommand{\calB}{\ensuremath{\mathcal{B}}\xspace}
\newcommand{\calL}{\ensuremath{\mathcal{L}}\xspace}
\newcommand{\calM}{\ensuremath{\mathcal{M}}\xspace}
\newcommand{\bfB}{\ensuremath{\mathbf{B}}\xspace}
\newcommand{\bfI}{\ensuremath{\mathbf{I}}\xspace}

\newcommand{\calI}{\text{$\mathcal{I}$}\xspace}

\newcommand{\norm}{\ensuremath{\Omega}\xspace}
\newcommand{\nnf}{\ensuremath{\mathsf{nnf}}\xspace}
\newcommand{\pos}{\ensuremath{\mathsf{pos}}\xspace}

\newcommand{\trans}{\ensuremath{\mathit{trans}}\xspace}
\newcommand{\ar}{\ensuremath{\mathit{ar}}\xspace}

\newcommand{\Prog}{\ensuremath{\Pi}\xspace}
\newcommand{\Rule}{\ensuremath{\rho}\xspace}

\newcommand{\impl}{\text{:$-$}\xspace}

\newcommand{\body}[1]{B(#1)}
\newcommand{\bodyp}[1]{B^+(#1)}
\newcommand{\bodyn}[1]{B^-(#1)}
\newcommand{\BU}{{\ensuremath{B_{\cal U}}}}
\newcommand{\commadots}[0]{,\ldots ,}
\newcommand{\derives}{\la}

\newcommand{\head}[1]{H(#1)}
\newcommand{\la}{\leftarrow}
\newcommand{\naf}{{\it not}\,}

\newcommand{\U}{{\ensuremath{\cal U}}}

\newcommand{\aggrcount}{\ensuremath{\mathit{\#count}}}

\newcommand{\thing}{\ensuremath{\concept{\top}}\xspace}

\newcommand{\grnd}{\ensuremath{\mathit{Gr}}}

\newcommand{\AS}{\mathcal{AS}}

\newcommand{\Clingo}{\ensuremath{\mathtt{Clingo}}\xspace}
\newcommand{\Wolpert}{\ensuremath{\mathtt{Wolpertinger}}\xspace}
\newcommand{\HermiT}{\ensuremath{\mathtt{HermiT}}\xspace}

\title{Bound Your Models!\\ How to Make OWL an ASP Modeling Language}

\author
{Sarah Alice Gaggl, Sebastian Rudolph, Lukas Schweizer}
\institute{Technische Universit\"at Dresden}

\begin{document}

\maketitle


\begin{abstract}
To exploit the Web Ontology Language OWL as an answer set programming (ASP) language, we introduce the notion of bounded model semantics, as an intuitive and computationally advantageous alternative to its classical semantics. We show that a translation into ASP allows for solving a wide range of bounded-model reasoning tasks, including satisfiability and axiom entailment but also novel ones such as model extraction and enumeration. Ultimately, our work facilitates harnessing advanced semantic web modeling environments for the logic programming community through an ``off-label use'' of OWL.
\end{abstract}

\begin{keywords}
Answer Set Programming, Bounded-Model Semantics, Semantic Web
\end{keywords}

\section{Introduction}
Answer set programming (ASP) is a powerful declarative language for knowledge representation and 
reasoning~\cite{BrewkaET11}. In ASP the knowledge is encoded in a set of logical rules and interpreted under the \emph{stable
model semantics}~\cite{Gelfond1988,GelfondL91}. Recent developements led to powerful systems e.g.\ dlv~\cite{dlv},
and gringo/clasp~\cite{gekakaosscsc11a}, to name some of them, which are capable to solve a large variety of problems~\cite{Gebser2012}.
  In particular, ASP has
shown to be well suited for big combinatorial search problems, as the dedicated solvers are specially designed to enumerate
all solutions~\cite{asp_competition2013}.

However, it has often been noted that, while being a powerful and versatile formalism, popularity and widespread adoption of logic programming in general and answer set programming in particular is hindered by the non-availability of user-friendly and scalable editing environments. 

On the other side, 
formalisms coming with a more elaborate modeling tool support -- most notably the Web Ontology Language OWL \cite{owl2-overview} -- are often preferred, even if the application scenario actually is of a constraint-satisfaction type which does not go well with OWL's standard semantics allowing for models of arbitrary size. Ontology editors like Prot\'{e}g\'{e} \cite{knublauch2004editing} provide user-friendly interfaces and combined with the natural language alike Manchester syntax \cite{owl2-man-syntax} possesses perspicuous access to a presumably complex and involved formalism.

We propose {\em bounded model reasoning} as an intuitive and simple approach to overcome this situation. Thereby, we endow OWL with a non-standard model-theoretic semantics and modifying the modelhood condition by restricting the domain to a finite set of bounded size, induced by the named individuals occurring in the given OWL ontology. We note that this additional condition can be axiomatized in the latest version of OWL. While reasoning in OWL under the classical semantics is \NTwoExpTime-complete  \cite{kazakov2008riq}, we show that reasoning under the {\em bounded model semantics} is merely \NP-complete. Still, employing the axiomatization, existing OWL reasoners struggle on bounded model reasoning, due to the heavy combinatorics involved.

Therefore, we propose a different approach and definine a translation of \sroiq knowledge bases (the logical counterparts to OWL ontologies) into answer set programs~\cite{BrewkaET11}, such that the set of bounded models coincides with the set of answer sets of the obtained program, allowing us to use existing answer set solvers (see~\cite{asp_competition2013} for an overview) for bounded model
reasoning. Next to the inferencing tasks typically used in semantic web technologies, this approach also allows for solving other, non-standard reasoning problems like model enumeration.

The benefits are manifolded, whereas in this work we particularly emphasize OWL as modeling language for typical constraint-satisfaction-type problems. The translation based approach can be seen as higher-level layer on top of the ASP language. Although we focus on the description logic \sroiq and its native DL syntax, other syntax specifications like the OWL~2 Manchester Syntax \cite{owl2-man-syntax} very well strive towards user-friendliness by means of natural language features.

We have implemented the proposed approach, for which first preliminary evaluations on typical con\-straint-satisfaction-type problems not only demonstrate feasibility, but also suggest significant improvement compared to the axiomatized approach using highly optimized OWL reasoners.

The article is organized as follows. In Section~\ref{sec:preliminaries} we introduce the necessary background on description 
logics and ASP. Then, in Section~\ref{sec:bounded-models} we define the \emph{bounded model semantics} and analyze their
complexity. The particular encoding of a \sroiq knowledge base into ASP is given in Section~\ref{sec:encoding}. A preliminary evaluation of the implemented system is summarized in Section~\ref{sec:evaluation}. Finally, we conclude in Section~\ref{sec:future-work} and discuss possible future directions. 

\section{Preliminaries}\label{sec:preliminaries}
In this section we provide the necessary background of description logics and answer set programming.

\subsection{Description Logics}

OWL~2~DL, the version of the Web Ontology Language we focus on, is defined based on description logics (DLs, \cite{dlhandbook,DBLP:conf/rweb/Rudolph11}). We briefly recap the description logic $\mathcal{SROIQ}$ (for details see \cite{HKS06-sroiq}). \label{def:syntax} Let $N_I$, $N_C$, and $N_R$ be finite, disjoint sets called \emph{individual names}, \emph{concept names} and \emph{role names} respectively. These atomic entities can be used to form complex ones as displayed in Table~\ref{tab:SROIQ}.
\newcommand{\tuplei}[1]{(#1)}
\begin{table}[t]
 \caption{Syntax and semantics of role and concept constructors in \SROIQ{},\label{tab:SROIQ} where $a_1, \ldots a_n$ denote individual names, $\rolS$ a role name, $\rolR$ a role expression and $\conC$ and $\conD$ concept expressions.}
\begin{center}
 \begin{tabular}{l l l }
 \hline
  Name & Syntax & Semantics \\\hline
  inverse role & $\rolS^-$ & $\{\tuplei{x,y}\in\Delta^\Inter\times\Delta^\Inter \mid \tuplei{y,x} \in \rolS^\Inter\}$ \\
  universal role & $\rolU$ & $\Delta^\Inter\times\Delta^\Inter$ \\
  top & $\top$ & $\Delta^\Inter $\\  
  bottom & $\bot$ & $\emptyset$ \\  
  negation & $\neg \conC$& $\Delta^\Inter \setminus \conC^{\Inter}$\\  
  conjunction & $\conC\sqcap \conD$& $\conC^{\Inter}\cap \conD^{\Inter}$ \\  
  disjunction & $\conC\sqcup \conD$& $\conC^{\Inter}\cup \conD^{\Inter}$\\  
  nominals   & $\{a_1,\ldots,a_n\}$ & $\{a_1^{\Inter},\ldots,a_n^{\Inter}\}$ \\
  univ. restriction & $\forall \rolR.\conC$ & $\{x \mid \forall y. \tuplei{x,y} \in \rolR^{\Inter} \to y\in \conC^{\Inter}\}$\\  
  exist. restriction & $\exists \rolR.\conC$ & $\{x \mid \exists y. \tuplei{x,y}\in\rolR^{\Inter} \wedge y\in \conC^{\Inter}\}$\\  
  $\Self$ concept & $\exists\rolR.\Self$ & $\{x \mid \tuplei{x,x}\in\rolR^{\Inter}\}$\\
  qualified number & $\atmost{n}\rolR.C$ & $\{x \mid \#\{y\in \conC^{\Inter}\mid \tuplei{x,y} \in \rolR^{\Inter}\}\le n\}$\\
  ~restriction & $\atleast{n}\rolR.C$ & $\{x \mid \#\{y\in \conC^{\Inter}\mid \tuplei{x,y} \in \rolR^{\Inter}\}\ge n\}$\\ \hline
 \end{tabular}
 \end{center}
\vspace{-2ex}
\end{table}
\begin{table}[t]
\caption{Syntax and semantics of $\mathcal{SROIQ}$ axioms.\label{tab:axm} }
\begin{center}
\begin{tabular}{ l l }
    \hline
 Axiom  $\alpha$ & $\mathcal{I}\models\alpha$, if \\ \hline \small
    $\rolR_1\circ\dots\circ \rolR_n\sqsubseteq \rolR$ \quad\quad& $\rolR_1^\mathcal{I}\circ\dots\circ \rolR_n^\mathcal{I}\subseteq \rolR^\mathcal{I}$ \hspace{10mm} \mbox{RBox }$\mathcal{R}$\\
    $\textsf{Dis}(\rolS,\rolR)$ & $\rolS^\mathcal{I} \cap \rolR^\mathcal{I}=\emptyset$ \\\hline
$C\sqsubseteq D$ & $C^\mathcal{I}\subseteq D^\mathcal{I}$\hfill \mbox{TBox }$\mathcal{T}$\\ \hline
$C(a)$ & $a^\mathcal{I}\in C^\mathcal{I}$\hfill \mbox{ABox }$\mathcal{A}$\\
$\rolR(a,b)$ & $(a^\mathcal{I},b^\mathcal{I})\in \rolR^\mathcal{I}$\\
$a\doteq b$ & $a^\mathcal{I}=b^\mathcal{I}$\\
$a\not\doteq b$ & $a^\mathcal{I}\neq b^\mathcal{I}$\\ \hline
  \end{tabular}
\end{center}
\end{table}
A \emph{$\mathcal{SROIQ}$ knowledge base} is a tuple $(\mathcal{A},\mathcal{T},\mathcal{R})$ where $\mathcal{A}$ is a $\mathcal{SROIQ}$ ABox, $\mathcal{T}$ is a $\mathcal{SROIQ}$ TBox and $\mathcal{R}$ is a $\mathcal{SROIQ}$ RBox. Table~\ref{tab:axm} presents the respective axiom types available in the three parts, and we will refer to each TBox axiom as {\em general concept inclusion} (GCI). The original definition of \sroiq contained more RBox axioms (expressing transitivity, (a)symmetry, (ir)reflexivity of roles), but these can be shown to be syntactic sugar. Moreover, the definition of \sroiq contains so-called \emph{global restrictions} which prevents certain axioms from occurring together. These complicated restrictions, while crucial for the decidability of classical reasoning in \sroiq are not necessary for the bounded-model reasoning considered here, hence we omit them for the sake of brevity.

The semantics of $\mathcal{SROIQ}$ is defined via interpretations $\mathcal{I}=(\Delta^\mathcal{I},\cdot^\mathcal{I})$ composed of a non-empty set $\Delta^\mathcal{I}$ called the \emph{domain of $\mathcal{I}$} and a function $\cdot^\mathcal{I}$ mapping individual names to elements of $\Delta^\mathcal{I}$, concept names to subsets of $\Delta^\mathcal{I}$ and role names to subsets of $\Delta^\mathcal{I}\times\Delta^\mathcal{I}$. This mapping is extended to complex role and concept expressions (cf. Table~\ref{tab:SROIQ}) and finally used to define satisfaction of axioms (see Table~\ref{tab:axm}). We say that  $\mathcal{I}$ satisfies a knowledge base $\KB=(\mathcal{A},\mathcal{T}, \mathcal{R})$ (or $\Inter$ is a model of $\KB$, written: $\mathcal{I}\models\KB$) if it satisfies all axioms of $\mathcal{A}$, $\mathcal{T}$, and $\mathcal{R}$. We say that a knowledge base $\KB$ \emph{entails} an axiom $\alpha$ (written $\KB\models \alpha$) if all models of $\KB$ are models of $\alpha$.

\subsection{Answer-Set Programming}
We give a brief overview of the syntax and semantics of disjunctive logic programs under the answer-sets semantics \cite{GelfondL91}.
We fix a countable set $\U$ of {\em (domain) elements}, also called \emph{constants}; and suppose a total order $<$ over the domain elements. An {\em atom} is an expression $p(t_{1},\ldots,t_{n})$, where $p$ is a {\em predicate} of arity $n\geq 0$ and each $t_{i}$ is either a variable or an element from $\U$.
An atom is \emph{ground} if it is free of variables. $\BU$ denotes the set of all ground atoms over $\U$. A \emph{(disjunctive) rule} \Rule is of the form
\begin{center}
\vspace{-2pt} $a_1\ \vee\ \cdots\ \vee\ a_n\ \la
        b_1,\ldots, b_k,\
        \naf b_{k+1},\ldots,\ \naf b_m$,
\end{center}
\vspace{-2pt} with $n\geq 0,$ $m\geq k\geq 0$, $n+m > 0$, where $a_1,\ldots ,a_n,b_1,\ldots ,b_m$ are atoms, or a \emph{count expression} of the form $\aggrcount\{l:l_1,\ldots,l_i\}\bowtie u$, where $l$ is an atom and $l_j = p_j$ or $l_j=\naf p_j$, for $p_j$ an atom, $1 \leq j \leq i$, $u$ a non-negative integer, and $\bowtie\; \in \{\leq,<,=,>,\geq\}$. Moreover, ``$\naf$'' denotes {\em default negation}. The \emph{head} of \Rule is the set $\head{\Rule}$ = $\{a_1\commadots a_n\}$ and the \emph{body} of \Rule is $\body{\Rule}= \{b_1,\ldots, b_k,\, \naf b_{k+1},\ldots,$ $\naf  b_m\}$. Furthermore, $\bodyp{\Rule}$ = $\{b_{1}\commadots b_{k}\}$ and $\bodyn{\Rule}$ = $\{b_{k+1}\commadots b_m\}$.
A rule \Rule is \emph{normal} if $n \leq 1$ and a 
\emph{constraint} if $n=0$. A rule \Rule is \emph{safe} if each variable in \Rule occurs in $\bodyp{r}$. A rule \Rule is \emph{ground} if no variable occurs in \Rule. A \emph{fact} is a ground rule with empty body and no disjunction. An \emph{(input) database} is a set of facts. A program is a finite set of rules. For a program \Prog and an input database $D$, we often write $\Prog(D)$ instead of $D\cup\Prog$. If each rule in a program is normal (resp.\ ground), we call the program normal (resp.\ ground).

For any program \Prog, let $U_\Prog$ be the set of all constants appearing in \Prog. $\grnd(\Prog)$ is  the set of rules $\Rule\sigma$ obtained by applying, to each rule $\Rule\in\Prog$, all possible substitutions $\sigma$ from the variables in \Rule to elements of $U_\Prog$. For count-expressions, $\{l:l_1,\ldots,l_n\}$ denotes the set of all ground instantiations of $l$, governed through $\{l_1,\ldots,l_n\}$.
%
An \emph{interpretation} $I\subseteq \BU$
\emph{satisfies} 
a ground rule \Rule iff $\head{\Rule} \cap I \neq \emptyset$ whenever $\bodyp{\Rule}\subseteq I$, $\bodyn{\Rule} \cap I = \emptyset$, and for each contained count-expression,   $N \bowtie u$ holds, where $N$ is the cardinality of the set of ground instantiations of $l$, $N = |\{l \mid l_1,\ldots,l_n\}|$, for $\bowtie\; \in \{\leq,<,=,>,\geq\}$ and $u$ a non-negative integer. $I$ satisfies a ground program $\Prog$, if each $\Rule\in\Prog$ is satisfied by $I$. A non-ground rule \Rule (resp., a program $\Prog$) is satisfied by an interpretation $I$ iff $I$ satisfies all groundings of \Rule (resp., $\grnd(\Prog)$). $I \subseteq \BU$ is an \emph{answer set} of $\Prog$ iff it is a subset-minimal set satisfying the \emph{Gelfond-Lifschitz reduct} $ \Prog^I=\{ \head{\Rule} \derives \bodyp{\Rule} \mid I\cap \bodyn{\Rule} = \emptyset, \Rule \in \grnd(\Prog)\} $. For a program $\Prog$, we denote the set of its answer sets by $\AS(\Prog)$.

\section{Bounded Models}\label{sec:bounded-models}

When reasoning in description logics, models can be of arbitrary cardinality. 
In many applications, however, the domain of interest is known to be finite. In fact, restricting DL reasoning to models of finite domain size (called \emph{finite model reasoning}, a natural assumption in database theory), has become the focus of intense studies lately \cite{Lutz2005,calvanese1996finite,rosati2008finite}.

As opposed to assuming the domain to be merely finite (but of arbitrary, unknown size), we consider the case where the domain has an \emph{a priori known cardinality}, more precisely, when the domain coincides with the set of named individuals mentioned in the knowledge base. We refer to such models as \emph{bounded models} and argue that in many applications this modification of the standard DL semantics represents a more intuitive definition of what is considered and expected as \emph{model} of some knowledge base.\footnote{In fact, when working with practicioners performing modeling tasks in OWL, we often found this to be their primary intuition, and OWL to be ``abused'' as a constraint language for an underlying fixed domain.}

\begin{definition}[Bounded-Model Semantics]\label{def:ib-interpretation}
Let \KB be a \sroiq knowledge base. An interpretation $\calI = (\Delta^\calI, \cdot^\calI)$ is said to be \emph{individual-bounded} \wrt \KB, if all of the following holds:
\begin{enumerate}
    \item $\Delta^\calI = \{ a \mid a \in \NiDL(\KB)\}$,\label{def:bounded-model-condition-1}
    \item for each individual $a \in \NiDL(\KB)$, $a^\calI = a$.\label{def:bounded-model-condition-2}
\end{enumerate}
Accordingly, we call an interpretation \calI \emph{(individual-)bounded model} of \KB, if \calI is an individual-bounded interpretation  \wrt \KB and $\calI \models \KB$ holds. A knowledge base $\KB$ is called \emph{bounded-model-satisfiable} if it has a bounded model. We say $\KB$ \emph{bounded-model-entails} an axiom $\alpha$ (written $\KB \models_\mathrm{bm} \alpha$) if every bounded model of $\KB$ is also a model of $\alpha$.
\end{definition}

\noindent Note that, under the bounded-model semantics, there is a one-to-one correspondence between (bounded) interpretations and sets of ground facts, if one assumes the set of domain elements fixed and known. That is, for every bounded-model interpretation $\calI = (\Delta^\calI,\cdot^\calI)$, we find exactly one Abox $\calA_\calI$ with atomic concept assertions and role assertions defined by $\calA_\calI := \{r(a,b) \mid (a,b) \in r^\calI\} \cup \{A(a) \mid a \in A^\calI\}$ and likewise, every such Abox $\calA$ gives rise to a corresponding interpretation $\calI_\calA$. This allows us to use ABoxes as representations of models.

We briefly demonstrate the effects of bounded model semantics as opposed to finite model semantics (with entailment $\models_\mathrm{fin}$) and the classical semantics.
Let $\KB=(\calA,\calT,\calR)$ with $\calA=\{A(a),A(b),s(a,b)\}$, $\calT=\{\top \sqsubseteq \exists r.B, \top \sqsubseteq \atmost{1}r^-.\top\}$, and $\calR=\{\textsf{Dis}(s,r)\}$.
First we note that $\KB$ has a bounded (hence finite) model $\calI$ representable as $\calA_\calI=\{A(a),A(b),B(a),B(b),s(a,b),r(a,a),r(b,b)\}$, thus $\KB$ is satisfiable under all three semantics.
Then $\alpha = \top\sqsubseteq \exists r.\exists r.B$ holds in all models of $\KB$, therefore $\KB \models \alpha$, $\KB \models_\mathrm{fin} \alpha$, and $\KB \models_\mathrm{bm} \alpha$. Opposed to this, $\beta = \top\sqsubseteq B$ merely holds in all finite models, whence $\KB \models_\mathrm{fin} \beta$ and $\KB \models_\mathrm{bm} \beta$, but $\KB \not\models \beta$. Finally, $\gamma = \top\sqsubseteq \exists r.\Self$ only holds in all bounded models, thus $\KB \models_\mathrm{bm} \gamma$, but $\KB \not\models_\mathrm{fin} \gamma$ and $\KB \not\models \gamma$.

\subsection{Extraction \& Enumeration of Bounded Models}\label{sec:bounded-ns-reasoning}
When performing satisfiability checking in DLs (the primary reasoning task considered there), a model constructed by a reasoner merely serves as witness to claim satisfiability, rather than an accessible artifact. However, as mentioned before, our approach aims at scenarios where a knowledge base is a formal problem description for which each model represents one solution. Then, retrieval of one, several, or all models is a natural task, as opposed to merely checking existence. With \emph{model extraction} we denote the task of materializing an identified model in order to be able to work with it, \ie to inspect it in full detail and reuse it in downstream processes.
The natural continuation of model extraction is to make all models explicit, performing \emph{model enumeration}. Conveniently, for both tasks we can use the introduced model representation via ABoxes. 

Most existing DL reasoning algorithms attempt to successively construct a model representation of a given knowledge base. However, most of the existing tableaux reasoners do not reveal the constructed model, besides the fact that in the non-bounded case models might end up being infinite such that an explicit representation is impossible. Regarding enumeration, we state that this task is not supported -- not even implicitly -- by any state-of-the-art DL reasoner, also due to the reason that in the non-bounded case, the number of models is typically infinite and even uncountable. We want to stick to the notions of model extraction and enumeration as their meaning should be quite intuitive. Although, in the general first-order case the term \emph{model expansion} is used, e.g.\ in the work of Mitchell and Ternovska \cite{mitchell2005framework}. There, an initial (partial) interpretation representing a problem instance is expanded to ultimately become a model for the encoded problem.

\subsection{Complexity of Bounded Model Reasoning}\label{sec:bounded-complexity}
The combined complexity of reasoning in \sroiq over arbitrary interpretations is known to be \NTwoExpTime-complete \cite{kazakov2008riq}. Still, it is considered to be usable in practice since worst-case knowledge bases would be of very artificial nature. Restricting to bounded models leads to a drastic drop in complexity.

\begin{theorem}\label{thm:bounded-complexity-npc}
The combined complexity of checking bounded-model satisfiability of \sroiq knowledge bases is \NP-complete.
\end{theorem}\vspace{-4mm}
\begin{proof}{(Sketch)}
To show membership, we note that after guessing an interpretation $\calI$, (bounded) modelhood can be checked in polynomial time.
For this we let $\mathcal{C}$ contain all the concept expressions occurring in $\mathcal{K}$ (including subexpressions). Furthermore, let $\mathcal{R}$ contain all role expressions and role chains (including subchains) occurring in $\mathcal{K}$. Obviously, $\calC$ and $\calR$ are of polynomial size. Then, in a bottom-up fashion, we can compute the extension $C^\calI$ of every element $C$ of $\calC$ and the extension $r^\calI$ of every element $r$ of $\calR$ along the defined semantics. Obviously, each such computation step requires only polynomial time. Finally, based on the computed extensions, every axiom of $\KB$ can be checked -- again in polynomial time.

To show hardness, we note that any 3SAT problem can be reduced to bounded-model satisfiability as follows:
Let $\calL=\{L_1,\ldots,L_n\}$ be a set of 3-clauses. Then satisfiability of $\bigwedge_{\{\ell_1,\ell_2,\ell_3\}\in \calL} (\ell_1\vee\ell_2\vee\ell_3)$ coincides with the bounded-model satisfiability of the knowledge base containing the two axioms $\top(a)$ and $\top \sqsubseteq \bigsqcap_{\{\ell_1,\ell_2,\ell_3\}\in \calL}(C_{\ell_1} \sqcup C_{\ell_2} \sqcup C_{\ell_3})$, where $C_{\ell_i}= A_p$ if $\ell_i = p$ and $C_{\ell_i}= \neg A_p$ if $\ell_i = \neg p$ for any propositional symbol $p$.
\end{proof}\noindent

Note that this finding contrasts with the observation that bounded-model reasoning in first-order logic is \PSpace-complete. We omit the full proof here, just noting that membership and hardness can be easily shown based on the fact that checking modelhood in FOL is known to be \PSpace-complete \cite{Stockmeyer:PhD} and, for the membership part, keeping in mind that \NPSpace=\PSpace{} thanks to Savitch's Theorem \cite{SavitchNPSpace}.
This emphasizes the fact that, while the bounded-model restriction turns reasoning in FOL decidable, restricting to \sroiq still gives a further advantage in terms of complexity (assuming $\PTime \not = \NP$).

\subsection{Axiomatization of Bounded Models inside \sroiq}\label{sec:bounded-tableaux}
When introducing a new semantics for some logic, it is worthwhile to ask if existing reasoners can be used.
Indeed, it is easy to see that, assuming $\{a_1,\ldots,a_n\}=\NiDL(\KB)$, adding the \sroiq GCI $\top \sqsubseteq \{a_1,\ldots,a_n\}$ as well as the set of inequality axioms containing $a_i \not \approx a_j$ with $i<j$ to $\KB$ will rule out exactly all the non-bounded models of $\KB$. Denoting these additional axioms with $\calB\calM$, we then find that $\KB$ is bounded-model satisfiable iff $\KB \cup \calB\calM$ is satisfiable under the classical DL semantics and, likewise, $\KB \models_\mathrm{bm} \alpha$ iff
$\KB\cup \calB\calM \models \alpha$ for any axiom $\alpha$. Consequently, any off-the-shelf \sroiq reasoner can be used for bounded-model reasoning, at least when it comes to the classical reasoning tasks.

However, the fact that the currently available DL reasoners are not optimized towards reasoning with axioms of the prescribed type (featuring disjunctions over potentially large sets of individuals) and that available reasoners do not support model extraction and model enumeration led us to develop an alternative computational approach based on ASP.


\section{Encoding \sroiq Knowledge Bases into ASP}\label{sec:encoding}

We propose an encoding of an arbitrary \sroiq knowledge base $\KB$, into an \emph{answer set program} $\Prog(\KB)$, such that the set of answer sets $\AS(\Prog(\KB))$, coincides with the set of bounded models of the given knowledge base. This allows us to use existing ASP machinery to perform both standard reasoning as well as \emph{model extraction} and \emph{model enumeration} quite elegantly. Intuitively, the set of all bounded models defines a search space, which can be traversed searching for models, guided by appropriate constraints. We thus propose an ASP encoding consisting of a generating part $\Prog_\mathrm{gen}{(\KB)}$, defining all potential candidate interpretations, and a constraining part $\Prog_\mathrm{chk}{(\KB)}$, ruling out interpretations violating the knowledge base.

Our translation into ASP requires a knowledge base in normal form which can be obtained by an easy syntactic transformation.


\begin{definition}[Normalized Form \cite{hermit}]\label{def:translation-normform}
A GCI is normalized, if it is of the form $\top \sqsubseteq \bigsqcup_{i=1}^{n} C_i$, where $C_i$ is of the form $B$, $\{a\}$, $\forall r.B$, $\exists r.Self$, $\neg \exists r.Self$, $\geq\,n\,r.B$, or $\leq\,n\,r.B$, for $B$ a literal concept, $r$ a role, and $n$ a positive integer. A TBox \calT\ is normalized, if each GCI in \calT\ is normalized. An ABox \calA\ is normalized if each concept assertion in \calA\ contains only a literal concept, each role assertion in \calA\ contains only an atomic role, and \calA\ contains at least one assertion.
An RBox \calR\ is normalized, if each role inclusion axiom is of the form $r \sqsubseteq r'$ or $r_1 \circ r_2 \sqsubseteq r'$.
A \sroiq\ knowledge base $\KB\!=\!(\calA,\calT,\calR)$ is normalized if \calA, \calT, and \calR\ are normalized.
\end{definition}

\begin{table}[t]
\caption{\label{tab:normalization}$\Omega$-Normalization of knowledge base axioms.}
$\begin{array}{@{}rll@{}}
\hline
\norm(\KB) &= &\displaystyle\bigcup_{\alpha\in\calR\cup\calA} \norm(\alpha) \cup \displaystyle\bigcup_{C_1\sqsubseteq C_2 \in \calT} \norm(\top \sqsubseteq \nnf(\neg C_1 \sqcup C_2)) \\
\norm(\top\sqsubseteq \mathbf{C} \sqcup C') &= &\norm(\top\sqsubseteq \mathbf{C} \sqcup \alpha_{C'}) \cup \displaystyle\bigcup_{1\leq i\leq n} \norm(\top\sqsubseteq \dot{\neg}\alpha_{C'} \sqcup C_i) \\
&& \text{for }C'\text{ of the form }C' = C_1 \sqcap \dots \sqcap C_n\text{ and }n \geq 2\vspace{2mm}\nonumber\\
\norm(\top\sqsubseteq \mathbf{C} \sqcup \forall r.D) &= &\norm(\top\sqsubseteq \mathbf{C} \sqcup\forall r.\alpha_D) \cup \norm(\top\sqsubseteq \dot{\neg}\alpha_D \sqcup D)\\
\norm(\top\sqsubseteq \mathbf{C}\; \sqcup \geq n\;r.D) &= &\norm(\top\sqsubseteq \mathbf{C}\, \sqcup \geq n\;r.\alpha_D) \cup \norm(\top\sqsubseteq\dot{\neg}\alpha_D\sqcup D)\\
\norm(\top\sqsubseteq \mathbf{C}\; \sqcup\leq n\;r.D) &= &\norm(\top\sqsubseteq \mathbf{C}\, \sqcup \leq n\;r.\dot{\neg}\alpha_{\dot{\neg}D}) \cup \norm(\top\sqsubseteq\dot{\neg}\alpha_{\dot{\neg}D} \sqcup \dot{\neg}D)\vspace{2mm}\\
\norm(\top\sqsubseteq \mathbf{C} \sqcup \neg\{s\}) &= &\left\{\begin{array}{ll}
\bot & \text{if }\mathbf{C} = \emptyset,\\
\norm(\mathbf{C}(s)) & \text{otherwise.}
\end{array}\right.\vspace{2mm}\\
\norm(D(s)) &= &\{\alpha_D(s)\} \cup \norm(\top\sqsubseteq\dot{\neg}\alpha_D \sqcup \nnf(D))\\
\norm(r^-(s,t)) &= &\{r(t,s)\}\\
\norm(r_1\circ\ldots\circ r_n\sqsubseteq r) &= & \{ r_1\circ r_2\sqsubseteq r_{(r_1\!\circ r_2)} \}\cup \norm(r_{(r_1\!\circ r_2)} \circ r_3\circ\ldots\circ r_n\sqsubseteq r)\\
& & \hfill%
\text{ for any RIA with $n>2$ }\\
\norm(\beta) &= &\{\beta\} \text{ for any other axiom }\beta\\
\hline \multicolumn{3}{c}{\alpha_C = \left\{
\begin{array}{ll}
Q_C & \text{if } \pos(C) = \true\\
\neg Q_C & \text{if } \pos(C) = \false
\end{array} \right., \text{where } Q_C \text{ is a fresh concept name unique for } C}\nonumber\\
\hline
\multicolumn{3}{@{}c@{}}{
\begin{array}{@{}c@{}}
\begin{array}{@{}rll@{}}
\pos(\top) &= &\false\\
\pos(A) &= &\true\\
\pos(\{s\}) &= &\true\\
\pos(\exists r.Self) &= &\true\\
\pos(C_1 \sqcap C_2) &= &\pos(C_1) \lor \pos(C_2)\\
\pos(\forall r.C_1) &= &\pos(C_1)\\
\pos(\geq n\; r.C_1) &= &\true
\end{array}
\begin{array}{rll@{}}
\pos(\bot) &= &\false\\
\pos(\neg A) &= &\false\\
\pos(\neg\{s\}) &= &\false\\
\pos(\neg\exists r.Self) &= &\false\\
\pos(C_1 \sqcup C_2) &= &\pos(C_1) \lor \pos(C_2)\\
\pos(\leq n\; r.C_1) &= &\left\{\begin{array}{ll}\pos(\dot{\neg}C_1) & \text{if }n=0\\\true & \text{otherwise}\end{array}\right.
\end{array}
\end{array}}\\
\hline \multicolumn{3}{l}{ \textbf{Note: } A \text{ is an atomic concept, } C_{(i)} \text{ are arbitrary concept expressions, } \bfC \text{ is a possibly}
}\nonumber\\
\multicolumn{3}{l}{\text{empty disjunction of concept expressions, }D \text{ is \textbf{not} a literal concept. The function }\dot{\neg}}\nonumber\\
\multicolumn{3}{l}{\text{is defined as }\dot{\neg}(\neg A) = A\text{ and }\dot{\neg}(A) = \neg A\text{ for some atomic concept }A.}\nonumber\\\hline\vspace{-9mm}\nonumber
\end{array}$
\end{table}

Given $\KB\!=\!(\calA, \calT, \calR)$, the normalized form $\norm(\KB)$ is obtained by applying a transformation $\norm$, given in Table \ref{tab:normalization}, which is mainly standard in DLs \cite{hermit}.
The normalized knowledge base $\norm(\KB)$ is a model-conservative extension of \KB, \ie every (bounded) model of $\norm(\KB)$ is a (bounded) model of $\KB$ and every (bounded) model of $\KB$ can be turned into a (bounded) model of $\norm(\KB)$ by finding appropriate interpretions for the concepts and roles introduced by $\Omega$. Thereby it is straightforward to extract a model for \KB, given a model of $\Omega(\KB)$. In the remainder, we will assume a knowledge base in normalized form, if not stated otherwise.

\subsection{Candidate Generation}
As shown, any potential bounded interpretation $\calI_\bfB$ is induced by a set of individual assertions \bfB, such that for each concept name $A$, role name $r$ and individuals $a,b$ occurring in \KB, either $A(a) \in \bfB$, or $\neg A(a) \in \bfB$ and either $r(a,b) \in \bfB$ or $\neg r(a,b) \in \bfB$. This construction is straightforward to encode via subsequent rules:\vspace{2mm}
\begin{small}\begin{align}
\Prog_\mathrm{gen}(\KB)    &:= &\{ \concept{A}(X)\ \impl\ \naf\ \concept{\neg A}(X),\, \concept{\thing}(X)\ |\ A \in \NcDL(\KB)\}\, \cup\label{eqn:translation-guess-rule-concept}\\
            &&\{\concept{\neg A}(X)\ \impl\ \naf\ \concept{A}(X),\, \concept{\thing}(X)\ |\ A \in \NcDL(\KB)\}\,\cup\vspace{2mm}\\
            &&\{ \ar(r,X,Y)\ \impl\ \naf\ \neg \ar(r,X,Y), \concept{\thing}(X), \concept{\thing}(Y)\ |\ r \in \NrDL(\KB)\}\,\cup\\
            &&\{ \neg \ar(r,X,Y)\ \impl\ \naf\ \ar(r,X,Y), \concept{\thing}(X), \concept{\thing}(Y)\ |\ r \in \NrDL(\KB) \}\,\cup\label{eqn:translation-guess-rule-role}\vspace{2mm}\\
            &&\{\concept{\top}(a)\,|\,a \in \NiDL(\KB)\}.\label{eqn:translation-guess-rule-thing}
\end{align}\end{small}\noindent
Recall, that a rule is \emph{unsafe}, if a variable that occurs in the head does not occur in any positive body literal. The predicate $\concept{\thing}(X)$ ensures safe rules, each of the guessing rules (\ref{eqn:translation-guess-rule-concept}--\ref{eqn:translation-guess-rule-role})  would otherwise be unsafe. This predicate  represents the $\top$-concept, to which the statement (\ref{eqn:translation-guess-rule-thing}) asserts each individual present in \KB. The function $\ar$ takes care of potential inverse roles (\cf Table \ref{tab:translation-naive-trbox}). Whereas ``$\naf$'' denotes default negation, $\neg$ is without attached semantics and merely used as syntactic counterpart to the DL vocabulary. We show now that $\Prog_\mathrm{gen}(\KB)$ computes $\calB_\KB$, the set of all constructible \bfB, and each $\bfB \in \calB_\KB$ determines a solution of $\Prog_\mathrm{gen}(\KB)$.

\begin{proposition}[$\calB_\KB = \AS(\Prog_\mathrm{gen}(\KB))$]\label{proposition:translation-generation} Let $\KB\!=\!(\calA, \calT, \calR)$ be a \sroiq knowledge base and $\Prog_\mathrm{gen}(\KB)$ the logic program obtained by the translation given in (\ref{eqn:translation-guess-rule-concept}--\ref{eqn:translation-guess-rule-thing}). Then, it holds that $\calB_\KB$ coincides with the set of all answer sets of $\Prog_\mathrm{gen}(\KB)$.
\end{proposition}

\subsection{Axiom Encoding}\label{sec:translation-axiom-encodings}

In the next step, we turn each axiom $\alpha \in \calR \cup \calT$ into a constraint, ultimately ruling out those candidate interpretations not satisfying $\alpha$. Moreover, each individual assertion in the ABox \calA restricts the search space further, since for some present fact $A(a)$ any solution candidate containing $\neg A(a)$ is eliminated. We will successively introduce appropriate encodings for axioms of each knowledge base component, altogether manifested in the program $\Prog_\mathrm{chk}(\KB)$ and will finally show that the program $\Prog(\KB) = \Prog_\mathrm{gen}(\KB) \cup \Prog_\mathrm{chk}(\KB)$ computes all bounded models of \KB.

\paragraph{Encoding TBox Axioms}
Since \calT is normalized, each GCI is of certain form which simplifies the encoding. We obtain $\Prog_\mathrm{chk}(\calT)$ as follows:
\begin{align}
\Prog_\mathrm{chk}(\calT)      &:=     &\{ \impl\ \trans(C_1), \ldots, \trans(C_n)\;|\;\text{for each } \top \sqsubseteq \bigsqcup^{n}_{i=1} C_i \text{ in } \calT\}\label{sec:translation-encoding-axiom-tbox}
\end{align}
Each concept expression $C_i$ is translated according to the function $\trans(C_i)$ depicted in Table \ref{tab:translation-naive-trbox}. Note, each $C_i$ is only one of the ones given in Definition \ref{def:translation-normform}, the ones given in the first column; \ie not complex, with the nice effect of $\trans(C_i)$ to be realized non-recursively.
\begin{table}
\caption{Translation of Concept Expressions.} \label{tab:translation-naive-trbox}
$\begin{array}{@{\hspace{2cm}}ll}
\hline
C               & \trans(C)\nonumber\\\hline
A               & \naf\ \concept{A}(X)\\
\neg A          & \concept{A}(X)\\
\{a\}               & \{\naf\ \concept{O_a}(X)\},\,\{\concept{O_a}(a)\}\\
\forall r.A         & \{ \naf\;\concept{A}(Y_A),\;\ar(r,X,Y_A) \}\\
\forall r.\neg A    & \{\ar(r,X,Y_A),\ \concept{A}(Y_A)\}\\
\exists r.Self      & \naf\;\ar(r,X,X)\\
\neg\exists r.Self  & \ar(r,X,X)\\
\geq n\ r.A         & \aggrcount\{ \ar(r,X,Y_A)\;:\; \concept{A}(Y_A)\} < n\\
\geq n\ r.\neg A    & \aggrcount\{ \ar(r,X,Y_A)\;:\; \naf\; \concept{A}(Y_A)\} < n\\
\leq n\ r.A         & \aggrcount\{ \ar(r,X,Y_A)\;:\;\concept{A}(Y_A)\} > n\\
\leq n\ r.\neg A\ \ \ & \aggrcount\{ \ar(r,X,Y_A)\;:\;\naf\;\concept{A}(Y_A)\} > n\\\hline \multicolumn{2}{l} { \textbf{Note: } O_a \text{ is a new concept name unique for a. And }\ar(r,X,Y) \text{ is defined as follows:}
}\nonumber\\
\ar(r,X,Y)          &:= \left\{\begin{array}{ll}\role{r}(X,Y) &\text{if $r$ is an atomic role}\\\role{s}(Y,X) &\text{if $r$ is an inverse role and $r = s^-$}\end{array}  \right.\nonumber\\\hline\vspace{-7mm}\nonumber
\end{array}$
\end{table}


\paragraph{Encoding RBox Axioms}
Role assertions and role inclusion axioms are also transformed into constraints, grouped in the program $\Prog_\mathrm{chk}(\calR)$. According to their DL semantics, this yields:
\begin{align}
\Prog_\mathrm{chk}(\calR)      :=\      &\{ \impl\ \ar(r,X,Y),\;\naf\;\ar(s,X,Y)\; |\;r \sqsubseteq s \in \calR\}\  \cup\\
                   &\{ \impl\ \ar(s,X,Y),\; \ar(r,X,Y)\; |\; \Dis(r,s) \in \calR\}\ \cup\label{sec:translation-encoding-axiom-dis}\\
                   &\{ \impl\ \ar(s_1,X,Y), \ar(s_2,Y,Z),\;\naf\; \ar(r,X,Z)\; | s_1 \circ s_2 \sqsubseteq r \in \calR\}. \label{eq:translation-chk-rbox-ria}
\end{align}\noindent
%

\paragraph{Encoding ABox Axioms}
The ABox \calA itself represents an input database, which we can directly use. However, it remains to check whether \calA does not contain contradictory knowledge; \ie propositional clashes of the form $\{A(a), \neg A(a)\} \in \calA$. Hence, the program $\Prog_\mathrm{chk}(\calA)$ consists of \calA and one additional constraint for each concept and role name ruling out inconsistent input ABoxes.
\begin{align}
\Prog_\mathrm{chk}(\calA)  :=\ &\calA\ \cup\\
                   &\{ \impl\ \concept{A}(X),\; \concept{\neg A}(X)\;|\;A \in \NcDL(\KB)\}\ \cup\\
                   &\{ \impl\ \ar(r,X,Y),\;\neg \ar(r,X,Y)\;|\;r \in \NrDL(\KB)\}.\label{sec:translation-encoding-axiom-role-consistency}
\end{align}
Note that the presence of $\{A(a), \neg A(a)\} \in \calA$ does not cause an unsatisfiable program under the answer set semantics, since $\neg$ does not have any meaning under the semantics; $\neg A$ is treated as just another predicate name. Thus, the imposed constraints simulate the known DL semantics.

\begin{theorem}\label{theorem:translation-theorem} Let $\KB\!=\!(\calA, \calT, \calR)$ be a normalized \sroiq knowledge base, and $\Prog(\KB)\!=\! \Prog_\mathrm{gen}(\KB) \cup\ \Prog_\mathrm{chk}(\KB)$ be the program obtained by applying Rules (\ref{eqn:translation-guess-rule-concept}--\ref{sec:translation-encoding-axiom-role-consistency}). Then, it holds:
\[
\AS(\Prog(\KB)) = \{\bfB\,|\,\bfB \in \calB_\KB \text{ and } \calI_\bfB \models \KB\}
\]
\end{theorem}



With this theorem in place, we benefit from the translation in many aspects. Most notably, in addition to the standard DL reasoning tasks, \emph{model extraction} and \emph{model enumeration} can be carried out without additional efforts, since both are natural tasks for answer set solvers.

\section{Evaluation}\label{sec:evaluation}

We implemented our approach as an open-source tool, named \Wolpert.\footnote{\url{https://github.com/wolpertinger-reasoner}} The obtained logic programs can be evaluated with most modern ASP solvers. However, the evaluation was conducted using \Clingo \cite{gekakaosscsc11a} for grounding and solving, since it currently is the most prominent solver leading the latest competitions \cite{asp_competition2013}. We present preliminary evaluation results based on simple ontologies, encoding constraint-satisfaction-type combinatorial problems. Existing OWL ontologies typically used for benchmarking, e.g.\ SNOMED or GALEN \cite{spackman1997snomed,rector1997experience}, do not fit our purpose, since they are modeled with the classical semantics in mind and often have little or no ABox information.

Our tests provide runtimes of \Wolpert and the popular \HermiT reasoner \cite{glimm2013hermit}. Whereas a direct comparison would not be fair, the conducted tests shall merely show the feasibility of our approach and the infeasibility of the axiomatization using standard DL reasoners. The evaluation itself is conducted on a standard desktop machine.\footnote{Unix operating system, $1.8$ Ghz Intel Core i7 Processor, $4\,$GB memory. Both tools are executed with standard Java-VM settings.}


\subsection{Unsatisfiability}

We construct an unsatisfiable knowledge base $\KB_n\!=\!(\calA_n, \calT_n, \emptyset)$, with $\calT_{\!n}$ and $\calA_n$ as follows:
\begin{align}
\calT_n     &= \{A_1 \sqsubseteq \exists r.A_{2},\dots,\; A_n \sqsubseteq \exists r.A_{n+1} \}\; \cup \\
        &\quad\ \{A_i \sqcap A_j \sqsubseteq \bot\;|\;1 \leq i < j \leq n+1\}\; 
        \vspace{2mm}\\
\calA_n     &= \{A_1(a_1),\top(a_1),\ldots,\top(a_n)\}
\end{align}\noindent
Inspired by common pigeonhole-type problems, we have $\KB_n$ enforce an $r$-chain of length $n+1$ without repeating elements, yet, given only $n$ individuals such a model cannot exist. Table \ref{tab:implementation-evaluation-unsat} depicts the runtimes for detecting unsatisfiability of $\KB_n$, for increasing $n$. The durations correspond to the pure solving time of \Clingo and pure reasoning time of \HermiT, respectively, as both \Wolpert and \HermiT have a comparable preprocessing. As the figures suggest, $\KB_n$ is a potential worst-case scenario, where both tools are doomed to test all combinations.
On this task, \Wolpert constantly outperforms \HermiT. For $\KB_{11}$, \HermiT is stopped after $15$ minutes, whereas \Wolpert detects unsatisfiability within $85$ seconds.

\begin{table}[t]
\caption{Runtime comparison for detecting unsatisfiability of $\KB_n$.}\vspace{-3ex}
\label{tab:implementation-evaluation-unsat}
\begin{center}
\begin{tabular}{ c  c  c  c  c  r }
\hline
\textbf{\#} & \textbf{Nominals} & \textbf{Concepts} & \textbf{Wolpertinger} & \textbf{HermiT} & \textbf{\#Backtrackings}\\ \hline
\textbf{1} &5 & 6 & $< 0,001\, $s & $000,148\,$s & $260$\\ 
\textbf{2} & 6 & 7 & $< 0,001\, $s & $000,436\,$s & $1.630$\\ 
\textbf{3} &7 & 8 & $000,040\,$s & $001,050\,$s & $11.742$\\ 
\textbf{4} & 8 & 9 & $000,320\,$s & $002,292\,$s & $95.900$\\ 
\textbf{5} & 9 & 10 & $004,790\,$s & $015,301\,$s & $876.808$\\ 
\textbf{6} &10 & 11 & $054,400\,$s & $144,024\,$s & $8.877.690$\\ 
\textbf{7} & 11 & 12 & $085,080\,$s & $> 15\,$min & $> 50 \times 10^6$\\ \hline
\end{tabular}\end{center}
\end{table}

\subsection{Model Extraction and Model Enumeration}
With Table \ref{tab:implementation-evaluation-enumeration-sudoku}, we next provide some figures for model extraction and partial enumeration (retrieving a given number of bounded models). To this end, we created a knowledge base modeling fully and correctly filled Sudokus, featuring $108$ named individuals, $13$ concept names and $1$ role name. When invoking a satisfiability test on this knowledge base using \HermiT, no answer was given within $15$ minutes.


On average, \Wolpert provides a solution for a given Sudoku instance in around $6$ seconds, of which more than $5$ seconds are needed for grounding, while the actual solving is done in less than $0.1$ seconds. For model enumeration, we used the knowledge base but removed information concerning pre-filled cells, turning the task into generating new Sudoku instances.
The size of the grounded program is $13\,$MB, the grounding process taking around $6$ seconds as reflected in Table \ref{tab:implementation-evaluation-enumeration-sudoku}.

\begin{table}[t]
\caption{Enumerating Sudoku Instances -- Runtime Overview.}\vspace{-3ex}
\label{tab:implementation-evaluation-enumeration-sudoku}
\begin{center}
\begin{tabular}{ c  r  r  r }
\hline
\textbf{\#} & \textbf{Models Requested} & \textbf{Time(Total)} & \textbf{Time(Solving)}\\ \hline
\textbf{1} & 100   & $006,014\,$s & $000,130\,$s \\
\textbf{2} & 1.000 & $006,372\,$s & $000,560\,$s \\ 
\textbf{3} & 10.000 & $008,558\,$s & $002,800\,$s \\
\textbf{4} & 100.000 & $034,096\,$s & $027,960\,$s \\
\textbf{5} & 1.000.000 & $279,059\,$s & $273,150\,$s \\
\hline
\end{tabular}\end{center}
\end{table}

\section{Conclusion}\label{sec:future-work}

With this paper, we have established the starting point for further developments on the theoretical and practical side, as well as we can identify benefits for both, the description logic and logic programming community. For the latter, our approach enables one to use OWL as ASP modeling language and therefore make use of the available tool support. Although modeling features are limited, we argue that quite large and involved problem scenarios can be modeled in OWL ontologies. Clearly, evaluations of our system with respect to such ontologies remain as imperative issue.

Complementarily, model extraction and enumeration supplement DL reasoning tasks for which our ASP translation represents not only a feasible approach, but apparently also a use case of ASP in another research field. Moreover, the framework may be extended to realize non-standard reasoning tasks useful for debugging purposes such as axiom pinpointing, explanation, justification and abduction, exploiting the innate capabilities of ASP to realize minimization as well as model enumeration.

On a more practical level, the proposed translation can certainly be optimized to exploit more built-in features of today's ASP solvers. 
In terms of harnessing the convenience of OWL modeling environments, we will implement an OWL API reasoner interface for \Wolpert, such that it can e.g., be seamlessly be integrated with other OWL software, such as Prot\'{e}g\'{e} \cite{knublauch2004editing}.

Regarding future theoretical DL investigations, in recent years, significant extensions of the modeling and querying capabilities of DLs have been proposed and partially implemented.
A major such extension is considering the reasoning task of answering queries, most prominently (unions of) conjunctive queries, positive queries,
conjunctive 2-way regular path queries, and monadically defined queries subsuming all of the former \cite{RK13:flagcheck}.
It is not overly difficult to show that answering all these query types over \sroiq knowledge bases (and hence over OWL ontologies) under the bounded model semantics is $\Pi^2_P$-complete, which again contrasts with the much worse results (if any) for the unbounded case \cite{RuGl10a,DBLP:conf/dlog/GlimmKL11}. Moreover, as all these query formalisms can be straightforwardly expressed in a rule-based way, an integration in our framework is immediate. In the same way, rule-based extensions of OWL -- monotonic \cite{SWRL,DLsafe} or nonmonotonic \cite{DBLP:journals/jacm/MotikR10,HybMKNF} -- should be straightforward to accommodate, at the cost of the combined complexity jumping to \ExpTime or \NExpTime.

\section*{Acknowledgements}
We are grateful for all the valuable feedback from our colleagues and the anonymous workshop reviewers, which helped greatly to improve this work.

\bibliographystyle{splncs03}
\bibliography{main-bib}

\newpage
\appendix
\section{Proofs}

\paragraph{Proof of Theorem \ref{theorem:translation-theorem}}

By Proposition \ref{proposition:translation-generation}, $\Prog_\mathrm{gen}(\KB)$ computes the set $\calB_\KB$. It remains to show, that $\Prog_\mathrm{chk}(\KB)$ obeys the bounded model semantics, and consequently excludes each $\bfB \in \calB_\KB$ not inducing a bounded model $\calI_\bfB$.

\paragraph*{$\AS(\Prog(\KB)) \subseteq \{\bfB\,|\,\bfB \in \calB_\KB \mathit{\ and\ } \calI_\bfB \models \KB\}$}
Let $\bfI \in \AS(\Prog(\KB))$ be an answer set of $\Prog(\KB)$. From Proposition \ref{proposition:translation-generation}, we know $\bfI \in \calB_\KB$. We show now that the interpretation $\calI_\bfI$ induced by \bfI is a bounded model of \KB, and therefore $\calI_\bfI \models \alpha$, for each axiom $\alpha \in \KB$. Then, let
\begin{itemize}
\item[] $\alpha \in \calR$: we distinguish role disjointness, and role inclusion axioms:
    \begin{itemize}
        \item[]\underline{$\alpha \in \Ra$}: Let $\alpha = \Dis(r,s) \in \Ra$, then by definition of $\Prog_\mathrm{chk}(\calR)$, there is a ground constraint $\rho_\alpha = \impl\ \role{s}(a,b),\,\role{r}(a,b)$ in $\grnd(\Prog_\mathrm{chk}(\calR))$, for all individuals $a,b \in \NiDL(\KB)$. Since \bfI is an answer set, $\{\role{s}(a,b),\,\role{r}(a,b)\} \not\in \bfI$. Consequently either $(a,b) \in \role{s}^{\calI_\bfI}$, or  $(a,b) \in \role{r}^{\calI_\bfI}$, hence $\calI_\bfI \models \Dis(r,s)$.
        \item[]\underline{$\alpha \in \Rh$}: then let $\alpha$ be of the form $s_1\, \circ\, s_2 \sqsubseteq r$, with $s_1,s_2,r \in \NrDL(\KB)$, and $\rho_\alpha = \impl\ \role{s_1}(a_1, a_2),\, \role{s_2}(a_2, a_3),\,\naf\,\role{r}(a_1, a_3)$ be the ground constraint  in $\grnd(\Prog_\mathrm{chk}(\calR))$. Since \bfI is an answer set, we have that, if $\role{s_1}(a_1, a_2)$ and $\role{s_2}(a_2, a_3) \in \bfI$ implies $\role{r}(a_1, a_3) \in \bfI$. And consequently $(a_1, a_2) \in \role{s_1}^{\calI_\bfI}$, $(a_2, a_3) \in \role{s_2}^{\calI_\bfI}$ and $(a_1, a_3) \in \role{r}^{\calI_\bfI}$, thus $\calI_\bfI \models s_1\, \circ s_2 \sqsubseteq r$.
    \end{itemize}
\item[] $\alpha \in \calT$: then $\alpha$ is normalized and of the form $\top \sqsubseteq \bigsqcup_{i=1}^n C_i$. In Rule (\ref{sec:translation-encoding-axiom-tbox}), $\alpha$ is turned into a constraint $\rho_\alpha = \impl\ \trans(C_1),\,\ldots,\,\trans(C_n)$  in $\Prog_\mathrm{chk}(\calT)$.
Since \bfI is an answer set, it does not violate any of the grounded instances of $\rho_\alpha$ in $\grnd(\Prog_\mathrm{chk}(\calT))$. Suppose now towards contradiction, $\calI_\bfI$ induced by \bfI does not satisfy $\alpha$, $\calI_\bfI \not\models \alpha$. Then, $\calI_\bfI \not\models C_i$, for all $1 \leq i \leq n$. However, since \bfI does not violate $\rho_\alpha$, in each of the ground instantiations of $\rho_\alpha$, there is exists a $\trans(C_i)$ which is not satisfied by \bfI, $1\leq i \leq n$. Then, $C_i$ is one of the expressions given in Definition \ref{def:translation-normform}, and we distinguish:
\begin{itemize}
    \item[]\underline{$C_i = A$}:  then $\trans(C_i) = \naf\; A(X)$, and $A(a) \in \bfI$ for any $a \in \NiDL(\KB)$. Consequently $a \in A^{\calI_\bfI} \neq \emptyset$, which contradicts the assumption $\calI_\bfI \not\models C_i$.
    \item[]\underline{$C_i = \neg A$}: then $\trans(C_i) = A(X)$, and $\neg A(a) \in \bfI$ for any $a \in \NiDL(\KB)$. Consequently $a \in (\neg A)^{\calI_\bfI} \neq \emptyset$, which contradicts the assumption $\calI_\bfI \not\models C_i$.
    \item[]\underline{$C_i = \{a\}$}: then $\trans(C_i) = \{\naf\,\concept{O_a}(X)\}$ and $\concept{O_a}(a)$, thus necessarily $\concept{O_a}(a) \in \bfI$. In order to not satisfy $\trans(C_i)$, $X = a$. Consequently we have $a \in O_a^{\calI_\bfI}$ with $O_a$ as nominal guard concept, and therefore $\{a\}^{\calI_\bfI} = \{a\}$, which contradicts the assumption.
    \item[]\underline{$C_i = \forall r.A$}: then $\trans(C_i) = \{r(X,Y_A),\,\naf\;A(Y_a)\}$, and $A(b) \in \bfI$ whenever $r(a,b) \in \bfI$. Consequently, $(a,b) \in r^{\calI_\bfI}$ implies $b \in A^{\calI_\bfI}$, which contradicts the assumption $\calI_\bfI \not\models C_i$.
    \item[]\underline{$C_i =\;\geq n\,r.A$}: then $\trans(C_i) = \aggrcount\{\role{r}(X,Y_\concept{A})\,:\,\concept{A}(Y_\concept{A})\} < n$, and more or equal than $n$, say $m$, atoms $\role{r}(a,b) \in \bfI$ and $\concept{A}(b) \in \bfI$. Consequently we find also $m$ pairs $(a,b) \in r^{\calI_\bfI}$ and $b \in A^{\calI_\bfI}$, which contradicts the assumption.
    \item[] All remaining cases can be treated analogously.
\end{itemize}
\item[] $\alpha \in \calA$: $\bfI$ satisfies $\Prog_\mathrm{chk}(\calA)$, in particular $\calA \subseteq \bfI$. Moreover, none of the imposed constraints in $\Prog_\mathrm{chk}(\calA)$ is violated, proving consistency of \calA, and therefore $\{\concept{A}(a),\,\concept{\neg A}(a)\} \not\in \bfI$ and $\{\role{r}(a,b),\,\role{\neg r}(a,b)\} \not\in \bfI$ for all concept names $A \in \NcDL(\KB)$, role names $r \in \NrDL(\KB)$ and individuals $a,b \in \NiDL(\KB)$.
\end{itemize}

\paragraph*{$\AS(\Prog(\KB)) \supseteq \{\bfB\,|\,\bfB \in \calB_\KB \mathit{\ and\ } \calI_\bfB \models \KB\}$} Let $\calI_\bfB$ be a bounded model of \KB, induced by some $\bfB \in \calB_\KB$. We show that \bfB is an answer set of $\Prog(\KB) = \Prog_\mathrm{gen}(\KB) \cup \Prog_\mathrm{chk}(\KB)$. From Proposition \ref{proposition:translation-generation} we know, that \bfB is an answer set of $\Prog_\mathrm{gen}(\KB)$, thus it remains to show that \bfB satisfies $\Prog_\mathrm{chk}(\KB)$ and therefore does not violate any of the imposed constraints.  Since $\calI_\bfB \models \alpha$, for each $\alpha \in \KB$, let
\begin{itemize}
    \item[]$\alpha \in \calR$: we distinguish again role disjointness and role chain axioms.  \begin{itemize}
        \item[]\underline{$\alpha \in \Ra$}: Let $\alpha = \Dis(r,s) \in \Ra$ and $\rho_\alpha = \role{r}(X,Y),\,\role{s}(X,Y)$, be the constraint according to Rule (\ref{sec:translation-encoding-axiom-dis}). Since $\calI_\bfB  \models \alpha$, for all $a,b \in \NiDL(\KB)$ we find, that $\{r(a,b), s(a,b)\} \not\in \bfB$, and consequently none of the grounded instances of $\rho_\alpha$ is violated by \bfB.
        \item[]\underline{$\alpha \in \Rh$}: then $\alpha$ is of the form $s_1 \circ s_2 \sqsubseteq r$, with $s_1,s_2,r \in \NrDL(\KB)$, and $\rho_\alpha = \impl\;\role{s_1}(X,Y_1), \role{s_2}(Y_n,Z),\naf\;\role{r}(X,Z)$ is the constraint according to Rule (\ref{eq:translation-chk-rbox-ria}). Since $\calI_\bfB \models \alpha$, for all $a_1, a_2, a_3 \in \NiDL(\KB)$ we have that if $s_1(a_1,a_2)$ and $s_n(a_2,a_3) \in \bfB$, then $r(a_1,a_3) \in \bfB$. Consequently, $\rho_\alpha$ is not violated by \bfB.
    \end{itemize}
    \item[]$\alpha \in \calT$: then $\alpha$ is normalized and of the form $\top \sqsubseteq \bigsqcup_{i=1}^n C_i$, which is satisfied by $\calI_\bfB$, \ifft $C_i^{\calI_\bfB} \neq \emptyset$ for some $1 \leq i \leq n$. Let $\rho_\alpha = \impl\ \trans(C_1),\,\ldots,\,\trans(C_n)$ be the constraint obtained from $\alpha$, applying Rule (\ref{sec:translation-encoding-axiom-tbox}). We need to show that \bfB does not violate the constraint. Let $C_i$ be the concept expression for which $C_i^{\calI_\bfB} \neq \emptyset$ holds, $1 \leq i \leq n$, and $C_i$ is one of the expressions given in Definition \ref{def:translation-normform}, in particular we have:
    \begin{itemize}
        \item[]\underline{$C_i = A$}: then $A^{\calI_\bfB} = \{a\,|\,\concept{A}(a) \in \bfB\}$. Consequently, for each of those $\concept{A}(a) \in \bfB$, $\trans(A) = \naf\;\concept{A}(X)$ is not satisfied.
        \item[]\underline{$C_i = \neg A$}: then $\neg A^{\calI_\bfB} = \{a\,|\,\concept{\neg A}(a) \in \bfB\}$. Consequently, for each of those $\concept{\neg A}(a) \in \bfB$, $\trans(\neg A) = \concept{A}(X)$ is not satisfied.
        %
        %
        \item[]\underline{$C_i = \exists r.\Self$}: then $(\exists r.\Self)^{\calI_\bfI} = \{a\,|\,\role{r}(a,a) \in \bfB\}$. Consequently, \\$\trans(\exists r.\Self) = \naf\,\role{r}(X,X)$ is not satisfied for those $\role{r}(a,a) \in \bfB$.
        \item[]\underline{$C_i = \leq r\,n.A$}: then $(\leq r\,n.A)^{\calI_\bfB} = \{a\,|\,\#\{\role{r}(a,b) \text{ and } \concept{A}(b) \in \bfB\} = m \leq n\}$. Consequently, $\trans(\leq r\,n.A) = \aggrcount\{ \role{r}(X,Y_A)\;:\;\concept{A}(Y_A)\} > n$, is not satisfied since there are $m$ such $\role{r}(a,b) \in \bfB$ with $\concept{A}(b) \in \bfB$.
        \item[] All remaining cases can be treated analogously.
    \end{itemize}
    \item[]$\alpha \in \calA$: then $\alpha \in \bfB$, as well as $\alpha \in \Prog_\mathrm{chk}(\calA)$, since by definition $\calA \in \Prog_\mathrm{chk}(\calA)$. In general, since $\calI_\bfB \models \KB$, \calA is consistent and therefore $\{\concept{A}(a), \concept{\neg A}(a)\} \not\in \bfB$, as well as $\{\role{r}(a,b), \role{\neg r}(a,b)\} \not \in \bfB$ for all concept names $A$, role names $r$ and individuals
    $a,b$.\hfill$\Box$
\end{itemize}

\end{document}